**Reproducible Workflow**
**Anirudh Prabhu[1] (prabha2@rpi.edu), Peter Fox[1] (pfox@cs.rpi.edu)**
**[1]Tetherless World Constellation, Rensselaer Polytechnic Institute, Troy, NY, USA.**
*Synonyms*
Reproducibility; Research Workflows; Computational Workflow

# 1. Introduction

"Science advances on a foundation of trusted discoveries, reproducing an experiment is one important approach that scientists use to gain confidence in their conclusions (McNutt 2014)."

# 2. Reproducibility vs Replicability: Various and Potentially Conflicting Definitions

Reproducibility has been extensively discussed in recent literature and been consistently mentioned as an important component of scientific research (Bechhofer et al. 2011; Stodden 2015; Plesser 2018; Goble 2016; Jasny et al. 2011; Miceli 2019; Peng 2009; Peng 2011). The first appearance of the phrase "reproducible research" in a scholarly publication appears in an invited paper presented at the 1992 meeting of the Society of Exploration Geophysics (Claerbour and Karrenbach 1992; Barba 2018). While there is widespread agreement on the importance of reproducibility along with the commonly used "R" terminology, e.g. Replicability, and Repeatability. There is some disagreement in the usage of these terms and their semantics: often their definitions are contradictory.

Broadly reproducibility and replicability have been defined along 4 different perspectives (Barba 2018; National Academies of Sciences, Engineering, and Medicine 2019):
1. There is no distinction between the terms and hence they can be used interchangeably (Bechhofer et. al 2011; Open Science Collaboration 2012; Stodden 2015). A few papers that do not distinguish between Reproducibility and Replication, define various levels of reproducibility. These levels are defined based on either the specific domain of research or the scientific approach (Stodden 2011, 2013; Goodman et. al 2016)
2. Reproducibility, referring to instances in which the similar results are obtained by a different team using a different experimental setup. For computational research, this focuses on an independent group obtaining similar outcomes using artifacts they develop independently (Plesser 2018; Goble 2016; Jasny et al. 2011; Association for Computing Machinery Version 1.0 2016; Drummond 2009).
3. Reproducibility, referring to the ability of an independent team of researchers to obtain consistent results (the same results) with the same experiment as the original study/experiment. For computational research, this implies using the same data resources, algorithms, source code, programming environment as the original study (Claerbout and Karrenbach 1992; Miceli 2019; Peng 2009; Association for Computing Machinery Version 1.1 2020; Walters 2013).
4. A Reproducibility Spectrum, which starts with a minimum standard for judging scientific claims and ends with full replication of the same results using the same data resources, algorithms and source code, programming environment. (Peng 2011)

Papers using approaches 2 and 3 have contradictory interpretations of the terms reproducibility and replicability. What approach 2 refers to as reproducibility is referred to as replicability in approach 3, and vice versa.

## 3. Is one definition needed?

With at least the four conflicting definitions, those cited in the previous section and many uncited in other review papers, the commonly asked question is: "it is possible to define reproducibility across domains and applications?". Varying answers are provided to this question, ranging from either picking a side in the conflicting definitions to just presenting an overview of the situation regarding terminology and its history (Barba 2018; Baker 2016; Plesser 2018). There does however exist some commonality between all of the definitions and interpretations of Reproducibility, Replicability, Repeatability and the other 'Rs'. The commonality observed in all the interpretations are **how** reproducibility and replicability (or however it is referred to across scientific literature) are achieved. Thus, instead of choosing one of the existing definitions, or creating a new one, focus would be on **how** to design and execute "a more complete" scientific experiment/study. While the Association for Computing Machinery's definitions (Association for Computing Machinery Version 1.0 2016; Version 1.1 2020) have taken one side and then the other in terms of defining reproducibility over the years, they also provide one aspect of introspection into the final task at hand. ACM appropriately divides the tasks for reproducibility or replicability (here after simply referred to as the 'Rs') as follows:

1. Same Team, Same Experimental Setup (STSE)
2. Different Team, Same Experimental Setup (DTSE)
3. Different Team, Different Experimental Setup. (DTDE)

If we focus on the non-functional requirements, the **how** of these 'Rs' rather the **what** they are called, then a more general consensus in science, at times even across domains is possible. That is because at the highest level, the 'Rs' will always fall into one of the 3 categories mentioned above. How researchers obtain consistent results for each of these 3 categorical conditions then becomes a way to distinctly define the 'Rs' for each approach, be it statistical, computational, empirical or experimental. For example, in STSE data-driven experiments (statistical and computational), one step includes setting a seed to reproduce results in methods that include randomization, especially statistical sampling or generating layouts of visualizations etc. And in DTSE setups, a common step is to store all intermediate results along with detailed documentation of any and all data processing that takes place in the experiments.

## 4. Workflows

Dramatic increases in the volume of data collected in the Geosciences in the last few decades have meant that there has been a trend involving the exploration of new data driven research methods. In the field of informatics, scientific explorations usually follow a set of steps to get from accessing the raw data, processing it, visualizing, analyzing and deriving insights, to answer scientific questions. Such a sequence of steps is called a workflow.

### 4.1. Workflow systems and the 'Rs'

Workflow systems (or Workflow Management Systems) have become increasingly popular as a way of specifying and executing data-intensive analysis (Davidson et al. 2008). A workflow system is a problem-solving environment, tuned to a distributed and service oriented computational infrastructure (Ludascher et al. 2006). Unlike general purpose scripting languages, workflow systems automatically record the means by which results are produced (Davidson et al. 2008).

During the initial development of the scientific workflow systems, the 'Rs' were not explicitly included or highlighted as part of the desiderata for workflow systems (Ludascher et al. 2006; McPhillips et al. 2009; Oinn et al. 2004; Hull et al. 2006). With the avalanche of data caused by the development of new experimental technologies, scientists increased their focus on data-driven and computational research methods (Cohen-Boulakia et al. 2017). This change in focus resulted in a renewed interest in the reproducibility (or replicability) of these experiments. Cohen-Boulakia et al. (2017) examined popular scientific workflow systems in the context of the 'Rs'. The authors do this by defining the various 'Rs' (what they call levels of reproducibility) and introducing a set of criteria that play a major role in the ability of a workflow systems to be "reproducibility-friendly" for a given computational or data-driven experiment. Popular workflow systems such as Taverna, Galaxy, OpenAlea, VisTrails and Nextflow were then examined in the context of reproducibility and their respective limitations were documented.

More recent platforms that help capture scientific workflows including the creation, execution, and publication of computational and data driven experiments, such as Whole Tale (Chard et al. 2020), consider and promote the 'Rs' as a central component of workflow design and execution.

## 5. 'Rs' in scientific workflows.

In reviewing extant literature and practice but without explicit statements confirming, we assert "While workflow systems have become popular recently, there is still significant portion of the scientific community that do not use workflow systems." In order to make sure data driven experiments achieve any one of more of the 'Rs', decisions need to be made during the workflow design process. This is especially important in case the workflows need to be reproduced (or replicated etc.) by researchers in different research domains (which is fairly common in interdisciplinary scientific explorations). Design decisions must made based on the non-functional requirements of the scientific experiments, i.e. based on **how** to design a reproducible (or replicable) workflow.

The categories mentioned in Peng's reproducibility spectrum (Peng 2011) are as follows (starting the least reproducible to the Gold Standard):
1. Publication + Code
2. Publication + Code + Data
3. Publication + Linked and executable code + Data
4. Full Replication (i.e. Gold Standard)

Since, in informatics, achieving the 'Rs' involve releasing data and code, we focus the next section on how to design workflows and share the required data and code to get closer to the gold standard.

As put forward by Sandve et al. (2013), "rules" for considerations in workflow design for ensuring the 'Rs' are fulfilled in the experiments are:
1. Keep track of how every result was produced.

2. Avoid Manual Data Manipulation.
3. Archive the exact versions of external programs used.
4. Version control all custom scripts.
5. Intermediate results need to be recorded, preferably in formats standardized throughout the experiment.
6. Note seeds set for analyses that include randomness.
7. Store raw data behind plots.
8. Generate detailed analysis outputs, allowing for inspections for summarization plots.
9. Connect text statements to underlying results.
10. Provide public access to scripts, runs and results.

Each of the above 10 rules capture a specific aspect of the 'Rs' and present what is needed in terms of handling the information and tracking procedures in order to release data and code for a scientific experiment (Sandve et al. 2013).

## 5.1. Releasing Data

Data used in an experiment/study can be obtained in a number of ways. It can range anywhere from being generated by the research team conducting the experiment/study, to being directly retrieved from an existing repository without making any changes to the data parameters. Usually, the process lies somewhere in the middle, where people need to retrieve the specific data required for their experiments from a few data repositories and make small additions or modifications to the data. In all of these cases though, there are some good practices to be followed so that the experiments achieve the criteria for the 'Rs'. These steps are as follows:

1. Credit the creators of the data resource(s) being used. This may include citing a data paper, a dataset PID or any citation/accreditation method stated by the dataset creators.
2. As the data is being processed, save versions of the dataset that are created during the execution of the workflow.
3. Document the code used for processing the data.

The datasets used in the experiment/study should be included along with the publication in many ways, such as:

- Releasing to data repositories like Zenodo, Dryad, Dataverse, Figshare etc. (Assante et al. 2016)
- Adding data files as part of the supplementary materials for the main publication. Some journals host data associated with their publications.
- Publishing data papers in scientific journals or the more nascent, but specific, data journals. (Callaghan et al. 2012).

## 5.2. Releasing Code

Releasing code is not only a central aspect of achieving the 'Rs' for an experiment/study, but also helps highlight the algorithms used, and decisions made throughout the research workflow. Similar to releasing data, code has been released in many ways:

- Adding the code/script files used in experiment along with the papers in the supplementary materials.
- Releasing code files and scripts as a single research product to repositories like Zenodo, Figshare etc. (Note: Unlike Dryand and Dataverse, Zenodo and Figshare are repositories that accept any research product, including code, presentations, videos, images, softwares etc.)

- Saving executable code for an experiment in an interactive environment like Jupyter or R notebooks, which are designed to support the workflow of scientific computing (Kluyver et al. 2016). Such notebooks can then be published to a repository for download and use or be linked to a working jupyterhub server. Services like mybinder.org enable sharing of projects containing live notebooks including a computational environment where user can execute code (Jupyter et al. 2018; Kluyver et al. 2016).

Writing code or scripts for an experiment/study are one of the steps where the research workflow is clearly visible. Hence, it is important to provide documentation with detailed instructions on how the individual code files tie into the larger research workflow. And also, correctly archive all the outputs (according to the rules put forward by Sandve et al. (2013)) required to achieve the 'Rs'.

## 6. Summary

In conclusion, even though there are some disagreements on defining reproducibility or replicability, we observe some consensus on the *how* to achieve reproducibility or replicability (the 'Rs'). Also, there has been progress in area of *data* and *code* releasing in the recent years which not only contributes to the 'Rs' of scientific research, but also provides researchers who work in these areas much needed credit and an incentive to share their work outside of a scientific journal paper.